\documentclass{article}
\usepackage{spconf,amsmath,graphicx}
\usepackage{color}
\usepackage{colortbl}
\usepackage{array}
\usepackage{graphicx}

\usepackage{enumitem}
\usepackage{stfloats}
\setlist{nosep, leftmargin=14pt}

\usepackage{mwe} % to get dummy images

\title{Polymerized Feature-based Domain Adaptation for Cervical Cancer Dose Map Prediction}
\name{Jie Zeng$^{1 *}$, Zeyu Han$^{1 *}$\thanks{* The first two authors contributed equally }, Xingchen Peng$^{2}$, Jianghong Xiao$^{3}$, Peng Wang$^{1}$, Yan Wang$^{1 \dagger}$\thanks{$\dagger$ Corresponding author: wangyanscu@hotmail.com}}

\address{$^{1}$ School of Computer Science, Sichuan University, China \\
    $^{2}$ Department of Biotherapy, Cancer Center, West China Hospital, Sichuan University, China \\
    $^{3}$  Department of Radiation Oncology, Cancer Center, West China Hospital, Sichuan University, China
    \vspace{-0.1em}
    }

\begin{document}
%\ninept
%
\maketitle
\begin{abstract}
  \vspace{-0.3em}
  Recently, deep learning (DL) has automated and accelerated the clinical radiation therapy (RT) planning significantly by predicting accurate dose maps. However, most DL-based dose map prediction methods are data-driven and not applicable for cervical cancer where only a small amount of data is available. To address this problem, this paper proposes to transfer the rich knowledge learned from another cancer, i.e., rectum cancer, which has the same scanning area and more clinically available data, to improve the dose map prediction performance for cervical cancer through domain adaptation. In order to close the congenital domain gap between the source (i.e., rectum cancer) and the target (i.e., cervical cancer) domains, we develop an effective Transformer-based polymerized feature module (PFM), which can generate an optimal polymerized feature distribution to smoothly align the two input distributions. Experimental results on two in-house clinical datasets demonstrate the superiority of the proposed method compared with state-of-the-art methods.
\end{abstract}
\section{Introduction}
\vspace{-1em}
\label{sec:intro}
% \cite{20, 21, 22, 23, 24, 25} [1-6]
As one of the most intractable cancers among women, cervical cancer is mainly treated by intensity-modulated radiation therapy (IMRT). Prior to conducting radiotherapy, dosimetrists need to incrementally adjust the input parameters of the treatment planning system in a trial-and-error manner to obtain a clinically acceptable dose distribution map, which is time-consuming and highly dependent on the expertise of the dosimetrists. Thus, it is desirable to automate the procedure of dose distribution prediction for improving the efficiency and quality of the radiotherapy plan.\\
\indent In recent years, with the flourishing of deep learning, many dose map prediction methods have been proposed and achieved great success with satisfactory performance [1-8]. For instance, Zhan et al. \cite{2} employed a Mc-GAN to perform dose prediction for rectum cancer. However, most existing deep learning-based methods are data-driven, while for cervical cancer, the clinical data is scarce due to its low morbidity, making the deep models exhibit poor performance in generalization. Luckily, domain adaptation (DA) is a good solution for alleviating the data scarcity problem. Considering rectum cancer shares the same CT scanning area and organs at risk with cervical cancer, we designate it as the source domain and perform DA for knowledge transferring. To enhance the alignment between the domains and capture long-range information, we suggest using a Transformer architecture as the underlying backbone instead of CNN. Building on the cross-attention mechanism \cite{11} in Transformer, Xu et al. \cite{12} designed a CDtrans to generate a polymerized domain using the cross-attention mechanism to help align features from the source and target domain. Despite its performant results in DA, the main drawback of CDtrans is that not all generated polymerized domains help transfer knowledge from the source domain to the target domain.\\
\indent In this paper, we propose a two-stage polymerized feature-based DA method to transfer knowledge from rectum cancer to cervical cancer for accurate cervical cancer dose map prediction. The overall model consists of two networks: 1) an aggregated network $Agg$ to align the data distributions of source and target domains; 2) an inference network $Infer$ initialized with $Agg$ to excavate the specific target domain features to generate more accurate dose maps. To bridge the gap between the source and target domains, we introduce a key component, polymerized feature module, into both $Agg$ and $Infer$ networks. Furthermore, in order to handle the above-mentioned problem in CDtrans, i.e., generating an aggregated domain that is more helpful for transferring knowledge, we design an adaptation bridge loss based on the argument that there exists an optimal polymerized domain along the "shortest geodesic path" between two domains \cite{13}. In addition, to guarantee the shared knowledge learned by $Agg$ wouldn't be forgotten in the $Infer$ network, we designate the $Agg$ as a teacher network to provide an additional distillation loss for $Infer$. Experimental results show that our model outperforms the state-of-the-art (SOTA) dose prediction models.

\begin{figure}[t]
\label{Fig.1}

  \begin{minipage}[t]{1.0\linewidth}
    \centering
    \centerline{\includegraphics[width=8.5cm]{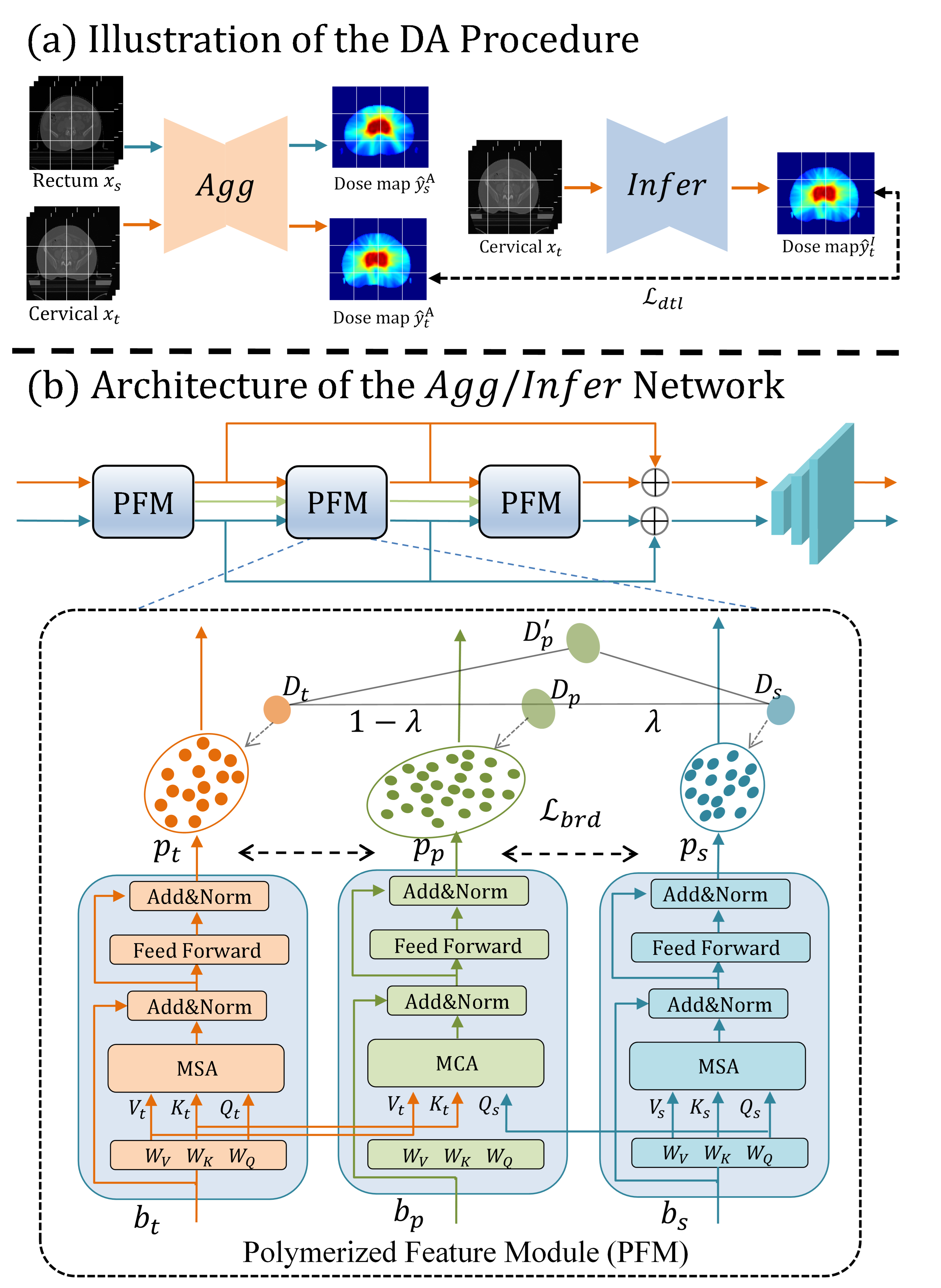}}
    \vspace{-1em}
    \caption{Overview of the proposed model. }
  \end{minipage}
  \vspace{-1.3em}
\end{figure}

\section{METHODOLOGY}
\vspace{-1em}
\label{sec:format}
The overall architecture of the proposed method is shown in Fig.1. Fig.1(a) demonstrates the entire DA procedure. The architecture of $Agg$ and $Infer$ is shown in Fig.1(b), where three PFM blocks are serially concatenated followed by a convolution network. The convolution network consists of two layers of Conv3×3, which is followed by BN and Relu, and a layer of Conv3×3 followed by Sigmoid. The details of the proposed method are described in the following subsections.

\vspace{-0.8em}
\subsection{Polymerized Feature Module}
\vspace{-0.6em}
\label{2.1}

\textbf{Weight-sharing Transformer Branches: }As shown in Fig.1(b), the PFM consists of three weight-sharing Transformer branches. Two of the branches namely $b_s$ and $b_t$ are used to learn domain-specific knowledge. The third branch namely polymerized branch $b_p$ is used to generate a polymerized feature distribution. The three branches have the same architecture with three learnable linear projectors $W_K$, $W_Q$ and $W_V$ to project the patches into three vectors, i.e., keys $K \in R^{N * d}$, queries $Q \in R^{N * d}$ and values $V \in R^{N * d}$, where $N$ is the number of patches and $d$ is the output dimension. For $b_s$ and $b_t$, we follow the traditional Transformer encoder to perform the multi-head self-attention (MSA). As for the polymerized branch $b_p$, compared with the other two branches, the only difference lies in the input. Specifically, the polymerized branch takes the queries $Q_s$ from the source domain, and keys $K_t$ and values $V_t$ from the target domain to conduct a multi-head cross-attention (MCA), which is formulated as follows:
\vspace{-0.8em}
\begin{equation}
\it{Att}_{cross} (Q_s,K_t,V_t )=softmax(\frac{Q_s K_t^T}{\sqrt{d}}) V_t.
\end{equation}
\par
% The cross-attention mechanism is capable of aggregating the input distributions of two different domains based on the similarity between them. Thus, even faced up with two extreme distributions, the generated intermediate distribution can serve as a kind of bridge to align them.
\vspace{-0.65em}
\textbf{Adaptation Bridge Loss: }The polymerized branch generates a polymerized feature distribution to help align two domains. However, not all polymerized distributions are appropriate for knowledge transfer. Here, we denote the distributions from source and target domains, and the polymerized distribution as $D_s$, $D_t$ and $D_p$ respectively. Assuming the different features located at different points in a manifold, there may be infinite polymerized features. Inspired by the "shortest geodesic path" \cite{13}, only those polymerized features located along the shortest geodesic path between the input distributions can be helpful for the smooth adaptation procedure. 
\par
Apparently, for a polymerized distribution $D_p^{\prime}$ which is not on the shortest geodesic path, the sum of $d(D_s,D_p^{\prime})$ and $d(D_t,D_p^{\prime})$ will be larger than the geodesic distance between two input distributions $d(D_s,D_t)$ according to the triangle inequality, where $d(\cdot )$ is the geodesic distance which is calculated by the $L_2$ distance between two distributions. As a result, this kind of polymerized distribution may lead to extra shifts, thus degrading the model performance. Hence, to guarantee the optimality of the polymerized feature, we employ a hyperparameter ${\lambda}{\in} {[ 0,1]}$ to balance the geodesic distance:
\begin{equation}
	\frac{d({D_s},{D_p} )}{d({D_t},{D_p} )} = \frac{\lambda}{(1-\lambda)}.
\end{equation}
\par
Further, derived from the above equation, we propose an adaptation bridge loss $L_{brd}$ to guarantee the optimality of the polymerized feature, which is expressed as follows:
\vspace{-0.8em}
\begin{equation}
L_{brd} = \lambda{\Vert p_s-p_p \Vert}_2+(1-\lambda) {\Vert p_t-p_p \Vert}_2,
\vspace{-0.8em}
\end{equation}
where $p_s$, $p_t$ and $p_p$ are the output representations from $b_s$, $b_t$ and $b_p$ respectively. Empirically, $\lambda$ is set to 0.5 to guarantee the impartiality of the polymerized feature distribution. By minimizing the $L_{brd}$ loss, the PFM can generate appropriate polymerized features, thus smoothing the feature alignment.
\vspace{-0.4em}
\subsection{Domain Adaptation Procedure}
\vspace{-0.6em}
\label{2.2}
Fig.1(a) illustrates the overall procedure of DA. To align the data distributions of two domains, we first train an aggregated network $Agg$ to learn the shared knowledge. Specifically, we input the original CT, PTV mask and OARs mask of the two domains into $Agg$ to generate the corresponding dose distribution maps. With three embedded PFMs, $Agg$ gradually minimizes the discrepancy between the two domains. To extract domain-specific knowledge, we employ $L_1$ loss to measure the difference between the predicted dose map and the ground truth, deriving a source domain loss $L_s$ and a target domain loss $L_t$. The formula can be expressed as follows:
\vspace{-0.8em}
\begin{equation}
  L_s = {\Vert y_s - \hat{y}_s^A \Vert }_1, \qquad   L_t = {\Vert y_t - \hat{y}_t^A \Vert }_1,
\vspace{-0.6em}
\end{equation}
where $(\hat{y}_s^A, y_s)$and $(\hat{y}_t^A, y_t)$ are predicted dose maps of the network $Agg$ and corresponding ground truth from the source domain and target domain, respectively. In summary, the overall loss of $Agg$ is calculated as follows:
\vspace{-0.8em}
\begin{equation}
  L_{A} = \sum_{i = 1}^{3} L_{brd}^{i} + \alpha L_s + \beta L_t,
  \vspace{-0.8em}
\end{equation}
where $i$ is the index of adaptation bridge loss calculated in the $i^{th}$ PFM in the aggregated model, $\alpha$ and $\beta$ are hyper-parameters for term balance.\par
% It worth noting that although the aggregated network $Agg$ can generate tolerable dose distribution maps, there are still tiny discrepancies between the predicted dose maps and the ground truth. The reason lies in the fact that some features only exist in the target domain but not in the source domain. 

% On the one hand, the $Infer$ network can be initialized with the weights learned from $Agg$ and fine-tuned in the target domain to accelerate the convergence of the model. On the other hand, the well-trained $Agg$ network can serve as a teacher to guide the learning of $Infer$, aiming at relieving the possible catastrophic forgetting problem. 
To further improve the quality of predicted dose maps, we introduce an inference network $Infer$ which only uses the target domain branch $b_t$ in the aggregated network to excavate the specific target domain features. The whole training process is that we pre-train $Agg$, then fix $Agg$ and train $Infer$. Specifically, we define a distillation loss $L_{dtl}$ to supervise the training of $Infer$:
\vspace{-1.2em}
\begin{equation}
  L_{dtl} = {\Vert \hat{y}_t^A - \hat{y}_t^I \Vert}_1,
  \vspace{-0.6em}
\end{equation}
where $\hat{y}_t^A$ and $\hat{y}_t^I$ are the predicted dose maps of the network $Agg$ and $Infer$, respectively. Designating $L_t^{\prime}$ as the $L_1$ loss between $\hat{y}_t^I$ and $y_t$, the overall loss of $Infer$ is formulated as:
\vspace{-1.2em}
\begin{equation}
  L_{I} = L_t^{\prime} + \gamma L_{dtl},
  \vspace{-0.6em}
\end{equation}
where $\gamma$ is the hyper-parameter for term balance.
\vspace{-0.4em}
\subsection{Training Settings}
\vspace{-0.6em}
The proposed method is implemented by the Pytorch framework using a NVIDIA GeForce RTX 3090 GPU with 24GB memory. Both $Agg$ and $Infer$ are trained by the Adam optimizer for 300 epochs with a batch size of 4. The learning rates of the two networks are initially set to 5E-4 and 1E-4 for the first 200 epochs, respectively, and then linearly decay to 0. As for the hyper-parameters in Eq.(6) and Eq.(8), based on our iterative trials, we set $\alpha$, $\beta$ and $\gamma$ to 0.4, 0.7 and 0.4, respectively.

\vspace{-0.6em}
\section{EXPERIMENTS AND RESULTS}
\vspace{-0.8em}
\label{sec:majhead}
\subsection{Dataset and Evaluation}
\vspace{-0.6em}
We conduct our experiments on two in-house datasets, i.e., a rectum dataset consisting of 130 subjects and a cervical dataset consisting of 42 subjects. The experiment is conducted in a 6-fold cross-validation manner. Four common metrics, including conformality index (CI), homogeneity index (HI), $D_x$ (minimum absorbed dose that covers a percentage volume $x$ of the PTV) and $V_x$ (percentage volume that receives a dose level of at least $x$), are employed for quantitative evaluation. Same as \cite{23}, we also use the average prediction error (APE) to measure the difference between the predicted dose map and the ground truth. For more intuitive comparison, we also depict the dose volume histogram (DVH) curves of the predicted dose map and the ground truth.

\begin{table}[!h]
  \vspace{-1.1em}
  \begin{minipage}[ht]{1.0\linewidth}
    \centering
\caption{APE values of the dosimetry metrics regarding PTV and OARs in comparison with ablation models.}
\label{tab:my-table}
\resizebox{\textwidth}{!}{%
\begin{tabular}{lllllll}
\hline
\multicolumn{1}{l|}{} & \multicolumn{4}{c|}{PTV} & \multicolumn{2}{c}{OARs} \\ \hline
\multicolumn{1}{l|}{} & \multicolumn{1}{c}{HI} & \multicolumn{1}{c}{CI} & \multicolumn{1}{c}{$D_{98}$} & \multicolumn{1}{c|}{$D_{95}$} & \multicolumn{1}{c}{$D_{mean}$} & \multicolumn{1}{c}{$V_{50}$} \\ \hline
\multicolumn{1}{l|}{(a)Backbone($Agg$)} & 0.046±0.029 & 0.031±0.025 & 0.045±0.041 & \multicolumn{1}{l|}{0.040±0.035} & 0.082±0.016 & 0.057±0.019 \\
\multicolumn{1}{l|}{(b)+MCA} & 0.046±0.016 & 0.040±0.016 & 0.039±0.012 & \multicolumn{1}{l|}{0.036±0.008} & 0.092±0.025 & 0.051±0.036 \\
\multicolumn{1}{l|}{(c)+$L_{brd}$} & 0.022±0.019 & 0.026±0.018 & 0.028±0.025 & \multicolumn{1}{l|}{0.021±0.016} & 0.090±0.044 & 0.044±0.038 \\
\multicolumn{1}{l|}{(*)Independent MSA} & 0.018±0.014 & 0.030±0.012 & 0.020±0.022 & \multicolumn{1}{l|}{0.009±0.006} & 0.040±0.032 & 0.045±0.028 \\
\multicolumn{1}{l|}{(d)Backbone($Infer$)+$L_1$} & 0.023±0.014 & 0.026±0.020 & 0.018±0.016 & \multicolumn{1}{l|}{0.014±0.009} & 0.040±0.039 & 0.048±0.055 \\
\multicolumn{1}{l|}{(e)+$L_{dtl}$} & \textbf{0.014±0.010} & \textbf{0.013±0.012} & \textbf{0.011±0.011} & \multicolumn{1}{l|}{\textbf{0.008±0.008}} & \textbf{0.021±0.017} & \textbf{0.037±0.040} \\ \hline
 &  &  &  &  &  & 
\end{tabular}%
}
  \end{minipage}
  \vspace{-2em}
  \end{table}

% \usepackage{graphicx}

\subsection{Ablation Experiment}
\vspace{-0.6em}
\label{ssec:subhead}

To demonstrate the effectiveness of each component of the proposed method, we break them up and reassemble them based on the backbone model. Specifically, for the aggregated network $Agg$, the network with only MSA serves as (a) backbone and then (b) incorporates the polymerized branch $b_p$ to perform MCA, and (c) further add the adaptation bridge loss $L_{brd}$. For the initialized inference network $Infer$, we (d) supervise it with only $L_1$ loss and then (e) add the distillation loss $L_{dtl}$. In addition, we further train an (*) independent inference network without initialization from $Agg$. \par
The quantitative results are shown in Table 1. As observed, comparing (a) and (b), the incorporation of the $b_p$ actually reduces the APE by 0.006 $D_{98}$ and 0.005 $V_{50}$, and slightly decreases the APE of HI, demonstrating the efficacy of the polymerized feature. Then, incorporated with $L_{brd}$, the model performs better in all the criteria, verifying the effectiveness of the $L_{brd}$ to generate better polymerized feature. In the fine-tuning stage, the $L_{dtl}$ can further improve the performance of the model in almost all the criteria. Notably, when $Infer$ was trained from scratch, the model performance decreased, suggesting that the dose map prediction performance of $Infer$ could be further improved by using the rectal cancer data as an assisted source.

\begin{table}[!h]
  \vspace{-1.2em}
  \begin{minipage}[ht]{1.0\linewidth}
  \centering
  \caption{APE values of the dosimetry metrics regarding PTV, OARs in comparison with SOTA methods. * means our method is significantly better than the compared method with $p<0.05$ via paired t-test.}
  \label{tab:my-table}
  \resizebox{\textwidth}{!}{%
  \begin{tabular}{lccccccc}
  \hline
  \multicolumn{1}{l|}{} & \multicolumn{4}{c|}{PTV} & \multicolumn{3}{c}{OARs} \\ \hline
  \multicolumn{1}{l|}{} & HI & CI & $D_{98}$ & \multicolumn{1}{c|}{$D_{95}$} & $D_{mean}$ & $V_{40}$ & $V_{50}$ \\ \hline
  \multicolumn{1}{l|}{Unet-GAN} & 0.035±0.012* & 0.050±0.041* & 0.043±0.034* & \multicolumn{1}{c|}{0.039±0.036*} & 0.048±0.043 & 0.009±0.013 & \textbf{0.014±0.022} \\
  \multicolumn{1}{l|}{DoseNet} & 0.015±0.012 & 0.088±0.030* & 0.015±0.010 & \multicolumn{1}{c|}{0.013±0.007} & 0.030±0.038 & 0.023±0.035 & 0.057±0.024 \\
  \multicolumn{1}{l|}{Unet} & 0.043±0.050 & 0.172±0.050* & 0.047±0.048* & \multicolumn{1}{c|}{0.021±0.022} & 0.329±0.250* & 0.039±0.015* & 0.027±0.021 \\
  \multicolumn{1}{l|}{DeepLabv3+} & \textbf{0.009±0.035} & 0.069±0.054* & 0.056±0.031* & \multicolumn{1}{c|}{0.055±0.029*} & 0.030±0.026 & 0.031±0.053 & 0.071±0.061 \\
  \multicolumn{1}{l|}{HDnet} & 0.027±0.007* & 0.015±0.010 & 0.018±0.010 & \multicolumn{1}{c|}{0.010±0.008} & 0.031±0.030 & 0.021±0.020 & 0.037±0.032 \\
  \multicolumn{1}{l|}{DPN} & 0.021±0.017 & 0.024±0.019 & 0.012±0.014 & \multicolumn{1}{c|}{0.010±0.018} & 0.054±0.026* & 0.011±0.006 & 0.046±0.005 \\ \hline
  \multicolumn{1}{l|}{\textbf{Proposed}} & 0.014±0.010 & \textbf{0.013±0.012} & \textbf{0.011±0.011} & \multicolumn{1}{c|}{\textbf{0.008±0.008}} & \textbf{0.021±0.017} & \textbf{0.009±0.005} & 0.037±0.030 \\ \hline
   & \multicolumn{1}{l}{} & \multicolumn{1}{l}{} & \multicolumn{1}{l}{} & \multicolumn{1}{l}{} & \multicolumn{1}{l}{} & \multicolumn{1}{l}{} & \multicolumn{1}{l}{}
  \end{tabular}%
  }
  \end{minipage}
  \vspace{-1.2em}
  \end{table}

  \vspace{-2em}
\subsection{Comparison with SOTA Methods}
\vspace{-0.6em}
\label{sssec:subsubhead}

To demonstrate the superiority of the proposed method, we compare our method with six SOTA dose map prediction methods, including Unet-GAN \cite{20}, DoseNet \cite{21}, Unet \cite{22}, DeepLabV3+ \cite{23}, HDnet \cite{24} and DPN \cite{26}. Notably, for a fair comparison, all the comparison methods were pre-trained on the source domain and then fine-tuned on the target domain. The quantitative APE results with respect to PTV and OARs are given in Table 2. It can be seen that our method achieves the best result in terms of CI, $D_{98}$, $D_{95}$, $D_{mean}$ and $V_{40}$, and the second best result in terms of HI. For the metric $V_{50}$, Unet-GAN and Unet outperform our method, but they are significantly worse than the proposed method in terms of all the other metrics. In addition, most p-values between the proposed and SOTA methods are less than 0.05, indicating a significant improvement brought by the proposed method. \par
In addition to quantitative results, more intuitive results are shown in Fig.2 and Fig.3. As shown in Fig.2, compared with other methods, our method generates a more authentic dose map that is closer to the ground truth proved by the error maps in the second line. In Fig.3, it can be seen that the DVH curves produced by the proposed method coincide with the ground truth the best compared with other SOTA methods.

\begin{figure}[t]
  \vspace{-0.8em}
  \begin{minipage}[t]{1.0\linewidth}
    \centering
    \centerline{\includegraphics[width=8.5cm]{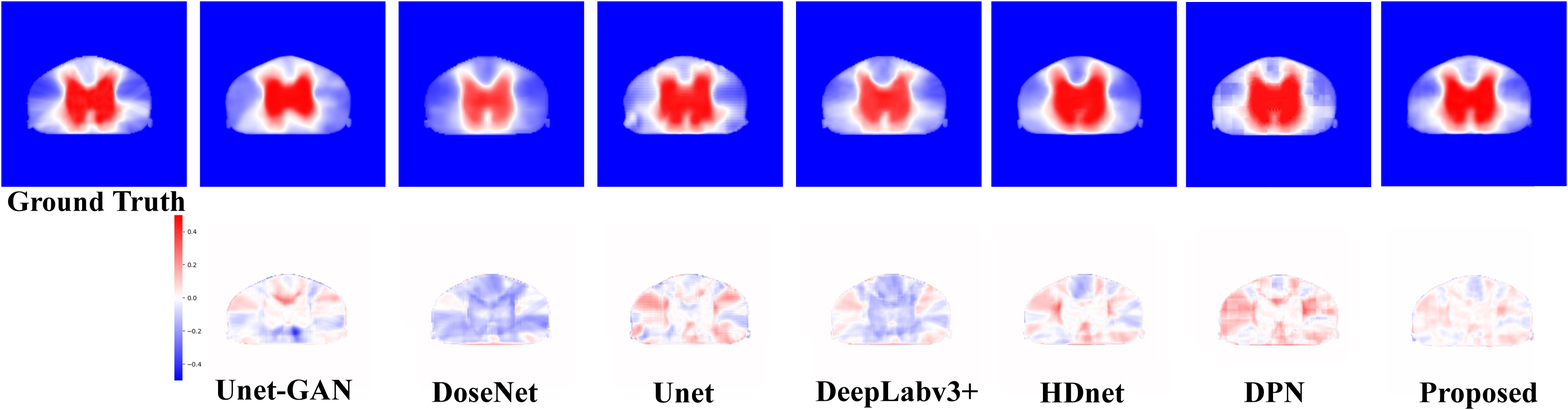}}
    \vspace{-2em}
    \caption{Visual comparison with SOTA models. First line is the ground truth and predicted dose maps, second line is the corresponding error maps.}
  \end{minipage}

\end{figure}

\begin{figure}[t]
  \vspace{-0.9em}
  \begin{minipage}[t]{1.0\linewidth}
    \centering
    \centerline{\includegraphics[width=8.5cm]{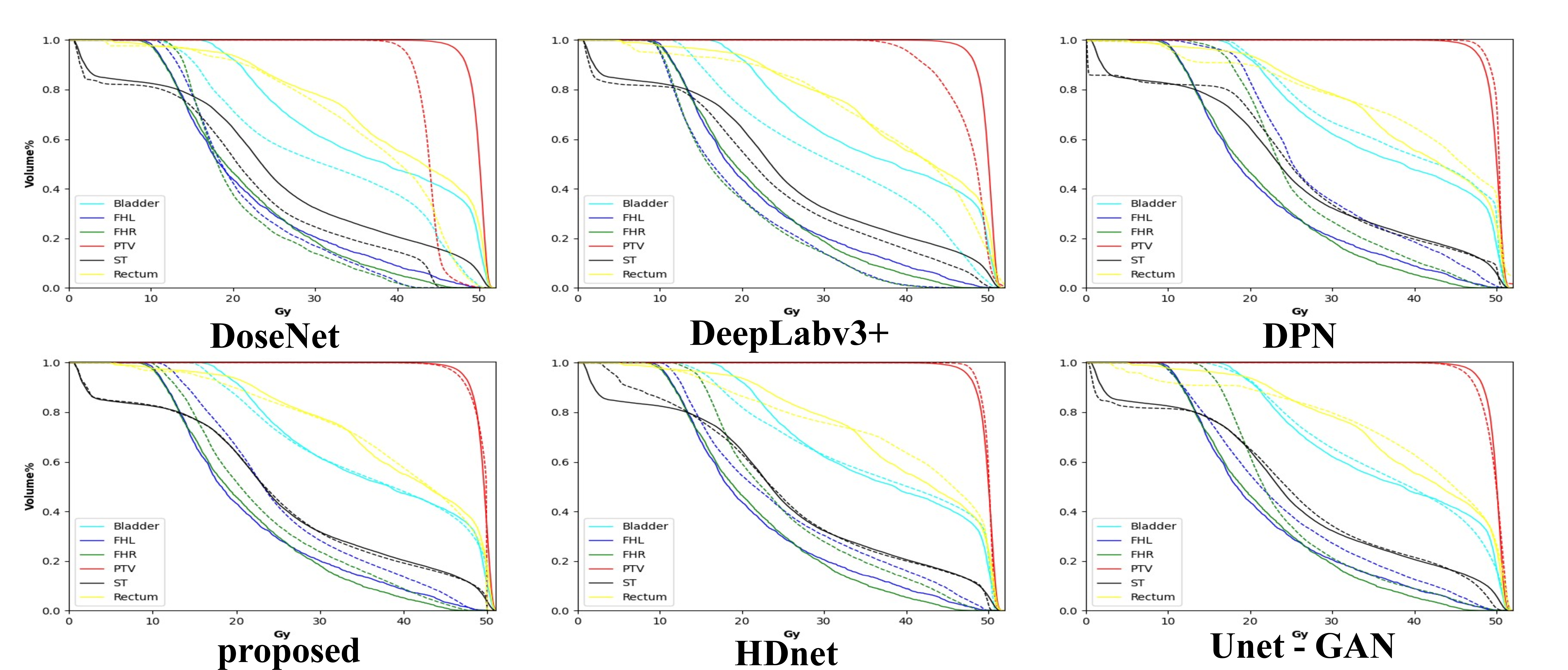}}
    \vspace{-1em}
    \caption{DVHs of real and predicted dose generated by different methods. Solid lines represent the original plan and dashed lines represent the predicted result.}
    \vspace{-1em}
  \end{minipage}
  \vspace{-0.5em}
\end{figure}

\vspace{-0.8em}
\section{CONCLUSION}
\vspace{-1em}
\label{sec:print}
In this paper, we propose a polymerized feature-based domain adaptation method to perform dose prediction for the radiotherapy of cervical cancer. The polymerized feature model (PFM) can obtain an optimal polymerized feature distribution with the help of adaptation bridge loss and then eventually fill the huge gap between two different distributions. After aligning the domain distributions, we fine-tune the network in the target domain with the distillation loss to preserve the well-learned shared knowledge. Experimental results demonstrate the feasibility and superiority of our method.
% \vfill
% \pagebreak
\vspace{-0.8em}
\section{Compliance with ethical standards}
\vspace{-1em}
\label{sec:ethics}

All procedures performed in studies involving human participants were in accordance with the ethical standards of the institutional and/or national research committee and with the 1964 Helsinki declaration and its later amendments or comparable ethical standards. This article does not contain any studies with animals performed by any of the authors.

\vspace{-0.8em}
\section{Acknowledgments}
\vspace{-1em}
\label{sec:acknowledgments}
This work is supported by National Natural Science Foundation of China (NSFC 62071314), Sichuan Science and Technology Program 2023YFG0263, 2023NSFSC0497, and Opening Foundation of Agile and Intelligent Computing Key Laboratory of Sichuan Province.

\end{document}